# THE EFFECTIVENESS OF APPLYING DIFFERENT STRATEGIES ON RECOGNITION AND RECALL TEXTUAL PASSWORD

Hassan Wasfi[1] and Richard Stone [2, 1]

[1]Iowa State University HCI Department, USA
[2]Iowa State University Industrial and Manufacturing Systems Engineering Department, USA

## ABSTRACT

*Using English words as passwords have been a popular topic in the last few years. The following article discusses a study to compare self-selection of the system-generated words for recognition and self-generated words for recall for nouns and mixture words. The results revealed no significant difference between recognition and recall of password nouns. The average memorability rate of noun recognition was 75.72%, slightly higher than noun recall 74.23% in long-term memory. Also, there was no significant difference between recognition and recall mixture word passwords. The average memorability rate of mixture word recognition was 95.23%, and recall was 84.14% in long-term memory. The authors concluded that the recognition and recall of mixed word passwords had a higher memorability rate than nouns.*

## 1. INTRODUCTION

Passwords provide a security mechanism to verify a users' identity before accessing private resources. A text-based password is one of the most popular methods to protect users' information [1, 2]. It is enduring ubiquitous because of the ease of password creation [3]. Humans tend to create an easy password containing their personal information, which is easy to be cracked [4, 5]. The users' behaviors of password creation enforce companies to establish password policies to increase security [6, 7, 8]. However, creating different passwords with different policies would be difficult to memorize [9, 10]. Several studies argued that utilizing passphrases instead of normal text passwords enhances security [11, 12, 13, 14]. Most studies are proposed as system-generated passphrases to produce long and unpredictable passwords [15, 16, 17]. The randomly generated words are difficult to remember because human memory is stimulated with the association between words. Also, the word types can play a role in storing a collection of words for long-term memory [18]. The passphrase structure significantly influences memorability, especially when built with no sequential or meaningful pattern. [19, 20, 21]. Therefore, auto-selection words negatively impact memorability, particularly when the words' types and structures are not equally distributed between each user [22]. Our study aimed to investigate the users' behaviors of choosing nouns or mixture words from a randomly generated set of words as a recognition password and compare it to a self-generated recall password.

                        

## 2. RELATED WORK

The forgettability issue of text-based passwords led researchers to discover an alternative aspect that enhances users' memorability. Engaging English words in the authentication field helps investigate its efficiency on human memory. The users' behaviors of creating a password from English words would differ from the normal text password. The majority of published papers are based on the passphrases approach to increase security [11, 15, 23]. Different techniques are occupied to randomly generated passphrases to enhance memorability, such as word associations and categorizations [24]. Also, occupying chunking theory decreases the load on user memory [25, 26]. However, passphrase has an issue with spelling mistakes because it includes more characters than traditional text passwords. Long passphrases raise the level of spelling mistakes [27, 28]. Several studies applied a typo-tolerant password protocol to avoid a small number of typographical errors of text-based passwords or phrases [29, 30, 31]. The system-generated passphrases have a problem related to word types distribution and structure [32, 33]. For instance, some participants commented that their passwords did not include a verb or nouns with a semantic meaning which influenced retrieving the correct password [19]. Moreover, the difficulty level of English words may influence memorizing a set of words, especially for non-native speakers [28]. Our study is proposed to analyze users' behaviors of password creation for recognition and recall textual password (RRTP):

- **Recognition:** selecting words from a set of randomly generated words presented on a grid.
- **Recall:** retrieving words created by users by typing in a textbox.

These concepts will be tested with two different types of English words (nouns and mixture words (nouns, verbs, adjectives, and helping words "is, are, and, but, in, has")). Our study will apply self-selection of system-generated words as recognition passwords and self-generated words as recall passwords. To reach to study goals, there are different hypotheses will be answered to have a clear idea of how participants will behave while creating their passwords:

**H1:** There will be a significant difference in memorability rate between self-selection of system-generated nouns as recognition passwords and self-generated nouns as a recall password.
**H2:** There will be a significant difference in login time between self-selection of system-generated nouns as recognition passwords and self-generated nouns as a recall password.
**H3:** There will be a significant difference in memorability rate between self-selection of system-generated mixture words as recognition password and self-generated mixture words as a recall password.
**H4:** There will be a significant difference in login time between self-selection of system-generated mixture words as recognition passwords and self-generated mixture words as a recall password.
**H5:** There will be a huge difference in the memorability rate between mixture words and nouns in long-term memory.

## 3. RESEARCH METHODOLOGY

This study will investigate the usability of RRTP for nouns and mixture words, as shown in figure1.This study proposed to analyze the users' behavior of self-selection of system-generated words as recognition passwords and compare it to self-generated words as a recall password. The majority of previous studies focused on a random selection method of recognition or recall passwords which have a negative impact on the memorability rate, as shown in table 1study (1, 2, 3). However, study (4) in table 1 stated that self-generated passphrase has a greater memorability rate than traditional passwords. Moreover, study (5) proved that self-generated passphrase has fewer cognitive load stressors on working memory than system-imposed passphrases. Our study is proposedto discover the structures of password creation for nouns and mixtures words and assess their role in memorability rate.

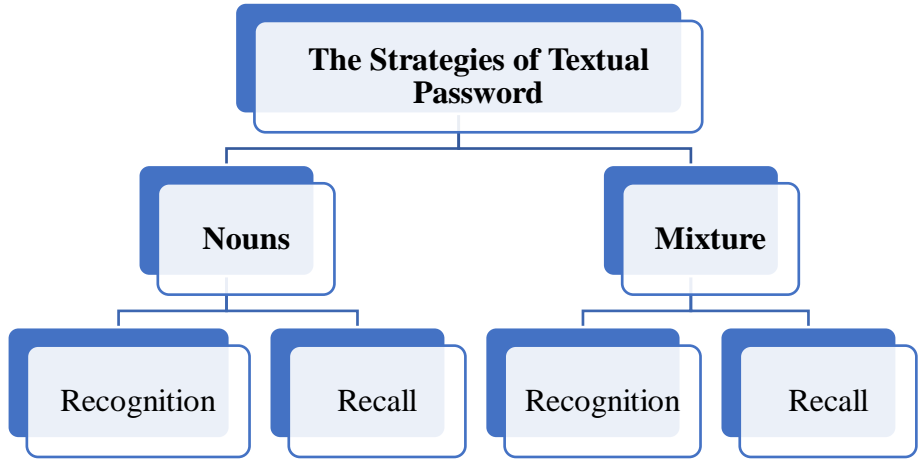

Figure 1. The strategies of textual password

Table 1. System and self-generated passphrase or nouns for different studies

| | Study | Methods | Words Type | Number of Words(passwords) | Comments |
|---|---|---|---|---|---|
| 1 | **Wright N, Patrick AS, Biddle R(2012)** | Recognition & Recall | System generated (nouns, verbs, adjectives) | 4 words | -Some participants commented that their password was not included a verb or semantic meaning **"throat"** and **"tongue"** which affected negatively retrieving the correct password. |
| 2 | **Assal H, Imran A, Chiasson S(2018)** | Recognition | System generated (nouns, verbs, adjectives) | 5 words | -No balance between word types presented to users in the registration phase. The words that appeared in GUI havefewer adjectives than other types. |
| 3 | **Shay R, Kelley PG, Komanduri S, Mazurek ML, Ur B, Vidas T, et al(2012)** | Recall | System generated nouns and passphrase | 4 words | -No analysis of how different types of nouns influence users' memorability.<br>- The successful login rate for system-generated nouns is |
| | | | | | slightly higher than different types of phrases. |
| 4 | **B. Bhana and S. V. Flowerday(2021)** | Recall | User-generated passphrase/password | Not stated | - passphrases users have a greater memorability rate than passwords which support the idea of long password lead to greater login failure. |
| 5 | **L. A. Loos, R. K. Minas, M.-B. C. Ogawa, and M. E. Crosby(2021)** | Recall | System and self-generated passphrase | 4 words | -Self-generated passphrases result in less cognitive load stressors on working memory than system-generated passphrases |

### 3.1. Participants

The total number of participants involved in this study was 74 (male = 41, female = 33). The participant's average age is 26 years, as shown in table 2. The majority of participants were a student at Iowa State University (93.24 %) and another part was from the public population (6.75 %).The education level of participants is undergraduate and higher. The participants were from different nationalities with native and non-native speakers.

Table 2. The number of participants and their age average for each group of RRTP

| Group | Male | Female | Average Age (Year) |
|---|---|---|---|
| **Group A (Nouns)** | 21 | 17 | 26 |
| **Group B (Mixture)** | 20 | 16 | 26 |

### 3.2. The Material

A mobile application was built to assess the user's memorability and log-in time for nouns and mixture words. Participant of recognition passwords has 24 words, randomly generated from a dictionary with 203 words. The nouns dictionary includes common nouns used in daily life. The mixture dictionary includes the common nouns, verbs, adjectives, and helping words ("is, are, but, and, in, has"). Each participant of the mixture recognition will have six nouns, six verbs, six adjectives, and six helping words. The theoretical space of recognition password is 18.43 bits which are less than required for security purposes, but this study will be focused on analyzing the user's behaviors. On the academic side, different studies were published on authentication systems with password space less than required for the system's usability purpose [34]. The theoretical space of the recall password is based on the number of words entered. The words set was taken from the Macmillan dictionary, and they were selected regarding the English level differences between participants because the study includes native and non-native speakers. The words were chosen based on our daily life use. Participants should select four words or more as recognition passwords by touching the computer screen. Users have to enter another four words or more from their mind as recall passwords by using a physical keyboard. A Dictionary checker was conducted to mitigate spelling mistakes.

## 3.3. Procedure

Before starting the study procedures, the participants were notified that the IRB of Iowa State University had approved our study. The users have to read through the consent form details and agree with the study procedures. The participants were informed that the password creation procedures would not include their private information or passwords. Each participant will make the experiment individually in a private lab. The participants were informed that the study would be three weeks long from the registration time as follow:

- **Short-Term Memory** (STM): After the registration time.
- **Long-Term Memory1**(LTM1): After one week of STM login.
- **Long-Term-Memory2**(LTM2): After two weeks of LTM1 login.

## 3.4. Pre-Survey

A pre-survey is required to specify the participants' age eligibility (over 18 years). Other information has been taken, such as gender, age, nationality, and education level, which would be helpful to analyze the relationship between these factors on the memorability rate.

## 3.5. Registration Phase

To investigate the memorability rate and users' habits of using English words as recognition and recall passwords, A user study was conducted. The user study was divided into two groups of English words:

- **Group A (within-subjects)** with two conditions recognition (DV) and recall (DV) of nouns, as shown in figure 2.

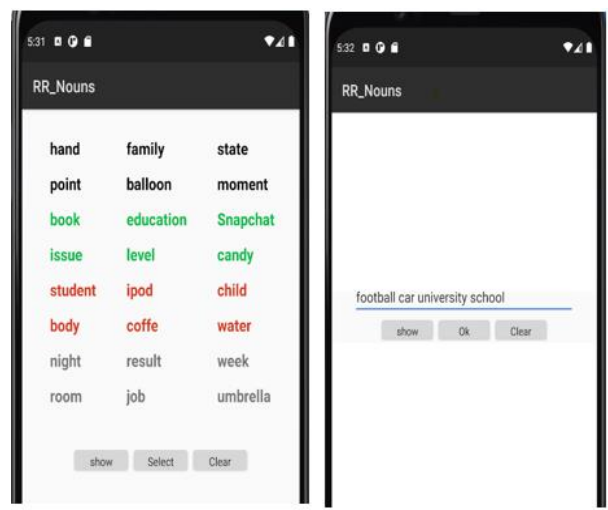

Figure 2. The recognition and recall passwords for nouns

- **Group B (within-subjects)** with two conditions recognition (DV) and recall (DV) of mixtures words (nouns, verbs, adjectives, and helping words ("is, are, but, and, in, has")), as shown in figure 3.

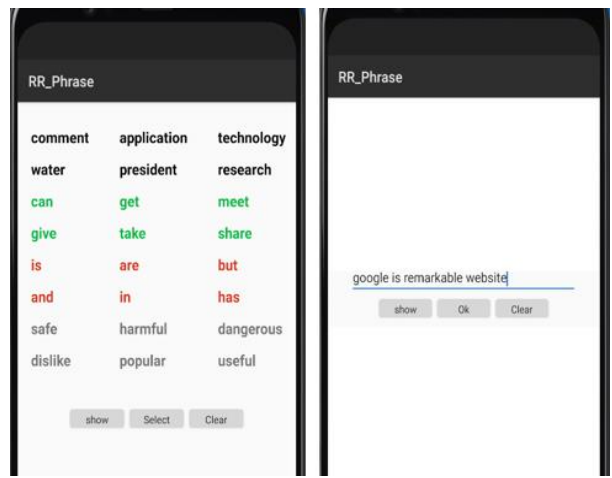

Figure 3. The recognition and recall passwords for mixture words

Each participant for both groups must follow the instructions below before creating the passwords:

- Not allowed to repeat any words of recognition password in the recall password.
- Not storing the words of password on paper or electronically.
- The typing words should be in a small letter.
- There will be a reminder for participants who failed to log in (in the lab).
- The mixture words participants should not use all words of their recognition or recall password from same words type. At least one or two words from different types.

### 3.6. Post-Survey

After the registration phase, the participants were asked to answer a post-survey to analyze the methods were used to build their recognition and recall passwords. Moreover, investigating the efficiency of those methods on their memory retrieval performance.

## 4. RESULT

### 4.1. Overview

The experiments are divided into two groups, as shown in table 3. The experiments are based on two different periods short-term and long-term memory. The memorability result of recognizing nouns is slightly higher than recall in LTM. The most factor that influenced users of nouns is the random selection such as **"story weather name Snapchat"**. The memorability result of mixture recognition is slightly higher than recall in LTM caused by users who did not grammatically structure their passwords, such as **"google become but dangerous"**. The experiments were in a private lab that helped control writing down or storing passwords electronically. The users who failed to log in have a reminder for their password in the lab.

Table 3. The study experiments of recognition and recall password

| | Study Experiments | | |
|---|---|---|---|
| | Memorability (STM) | Memorability (LTM 1 and 2) | User Behavior |
| **Experiment 1** | recall nouns> recognition nouns | recognition nouns > recall nouns | -one chunk<br>-two chunks<br>-scenario<br>-random |
| **Experiment2** | recognition mixture > recall mixture | recognition mixture > recall mixture | -grammar flow<br>-no grammar flow |
| **Result of Experiment 1 & 2** | The mixture password has a higher memorability rate compared to nouns for both conditions | | |

To assess the data normality, SPSS software applied skewness and kurtosis measurements across the noun's recognition and recall (STM, LTM1, and LTM2). Due to the confirmed non-normality of the data. The non-parametric Mann-Whitney test was applied. Also, to assess the data normality, SPSS software applied skewness and kurtosis measurements across the mixture recognition and recall (STM, LTM1, and LTM2). Due to the confirmed non-normality of the data. The non-parametric Mann-Whitney test was applied. The password entropy for each group is shown in table 4.

Table 4. The password entropy for recognition and recall of nouns and mixture words

| | Password Entropy | |
|---|---|---|
| | Recognition | Recall |
| **Group A** | 18.34 bits | 61.11< bits <178.62 |
| **Group B** | 18.34 bits | 61.11< bits < 136.31 |

## 4.2. The Memorability Rate of RRTP for Group A (Nouns)

**H1:** There will be a significant difference in memorability rate between user selection of system-generated nouns as recognition passwords and self-generated nouns as a recall password?

**H2:** There will be a significant difference in login time between user selection of system-generated nouns as recognition passwords and self-generated nouns as a recall password?

A Mann-Whitney test was conducted to test the significant difference in memorability rate between user selection of system-generated nouns (recognition) and self-generated nouns (recall). There was no significant difference in the memorability rate and login time between recognition and recall passwords, as shown in table 5. In STM, the memorability rate for recall password (mean= 41.63) was higher than recognition password (mean= 35.37), (p= .082).

Table 5. The memorability rate and login time of RRTP for nouns

| | | Memorability Rate and Login Time for Nouns Password | | |
|---|---|---|---|---|
| | | N | Memorability P-value | Login Time P-value |
| **STM** | **Recognition && Recall** | 38 | .082 | .010 |
| **LTM 1** | **Recognition && Recall** | 34 | .807 | .087 |
| **LTM 2** | **Recognition && Recall** | 33 | .806 | .521 |

In contrast, the recognition password has a slightly higher memorability rate than the recall password in LTM1 caused by a sharp reduction in recall memorability rate to 61.76 % compared to STM, as shown in figure 4. In LTM 2, the memorability rate increases compared to LTM 1, which belongs to 73.68 % of participants who used a reminder retrieved their password correctly, as shown in table 6. The recall password has less login time compared to recognition password in STM and LTM, as shown in figure 5.

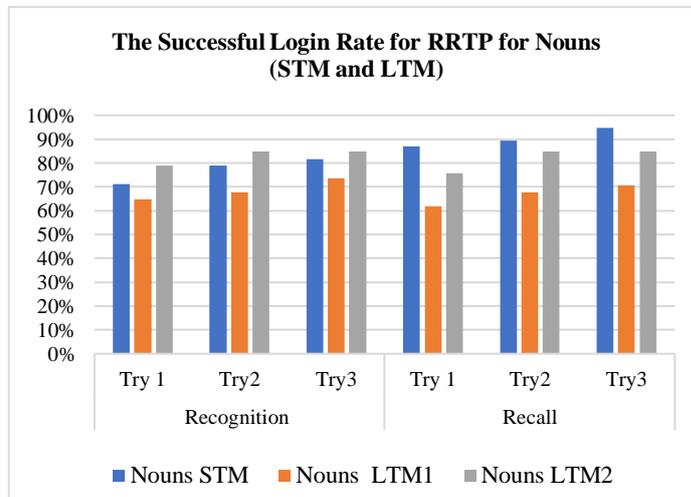

Figure 4. The successful login rate of RRTP for nouns

Table 6. The enhancement on memorability rate after a reminder nouns password

| Reminder (Nouns) | | | |
|---|---|---|---|
| Recognition | | Recall | |
| No. Participants | Correctness | No. Participants | Correctness |
| First Reminder (9) | 7 | First Reminder (2) | 0 |
| Second Reminder (9) | 7 | Second Reminder (10) | 7 |

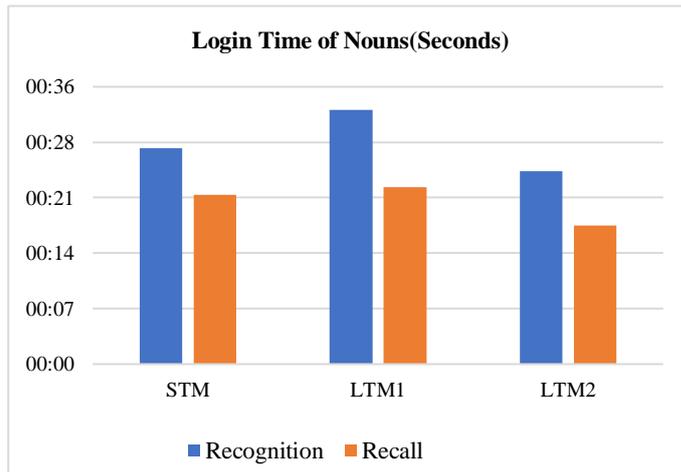

Figure 5. Login time for RRTP of nouns

The users have created their passwords based on four different structures: one chunk, two chunks, scenario, and random as shown in figure 6 and example for their structures, as shown in table 7.

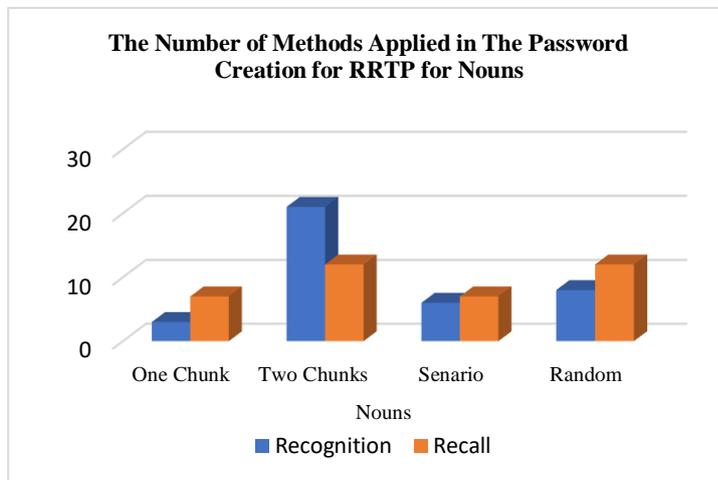

Figure 6. The methods applied for RRTP for nouns

Table 7. Examples of the methods used by participants for password creation for nouns

| Password Creation Methods Examples | | | |
|---|---|---|---|
| **One Chuck** | **Two Chunks** | **Scenario** | **Random** |
| -Same first letter (weather water world work)  -Family members (father mother brother sister)  -Information technology (computer google facebook iPod) | -gender+ life experience (mother woman life education)  -body parts+ sides (head eye back end) | -the relationship between education and work (education reason work life) | -no relationship (work party name business) (story weather name snapchat) |

The random method of recognition and recall password had a 57.5 % successful login average rate but one chunk (100 %), two chunks (86.27 %), scenario (86.05 %), as shown in figure 7.The reduction in the memorability rate was caused by either missing order or substituting with other words of their password. Therefore, one chunk, two chunks, and scenario methods proved a higher memorability rate than random selection for both conditions.

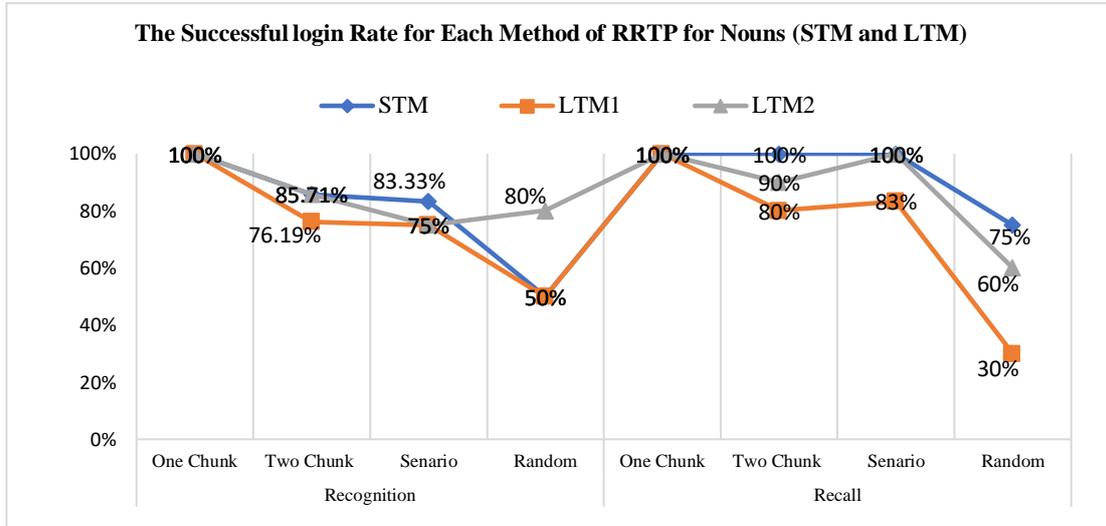

Figure 7. The successful login rate for each method of RRTP for nouns

### 4.3. The Memorability Rate of RRTP for Group B (Mixture)

**H3:** There will be a significant difference in memorability rate between user selection of system-generated mixture words as recognition passwords and self-generated mixture words as a recall password?

**H4:** There will be a significant difference in login time between user selection of system-generated mixture words as recognition passwords and self-generated mixture words as a recall password?

A Mann-Whitney test was conducted to test the significant difference in memorability rate between user selection of system-generated mixture words (recognition) and self-generated mixture words (recall). There was no significant difference in the memorability rate and login time between recognition and recall passwords in STM and LTM, as shown in table 8.

Table 8. The memorability rate and login time of RRTP for mixture word

|  |  | Memorability Rate and Login Time for Mixture Password | | |
|---|---|---|---|---|
|  |  | N | Memorability P-value | Login Time P-value |
| STM | Recognition && Recall | 35 | .626 | .147 |
| LTM 1 | Recognition && Recall | 32 | .213 | .060 |
| LTM 2 | Recognition && Recall | 28 | .079 | .153 |

The memorability rate for recognition passwords was higher than recall in STM and LTM, as shown in figure 8. The login time for both recognition and recall passwords are almost equal to each other, as shown in figure 9. The users of mixture words built more than 25 structures as a phrase for both conditions. Some of these structures had a negative impact in retrieving the correct password because they are not structured the words of their password in a meaningful format or grammatically correct, as shown in table 9. There are 8.57 % of participants had missed orders or substituted with other words of their password as well as, 5.71 % had a spelling mistake such as using plural instead of singular or name mistake using **"columbus"** instead of **"colombus"** although using spelling checker. For these mistakes, a reminder was used twice for recognition password and five times for recall password with 75 % enhancement in retrieving the correct password, as shown in table 10. Therefore, the mixture words password for both conditions have a stable memorability rate in LTM.

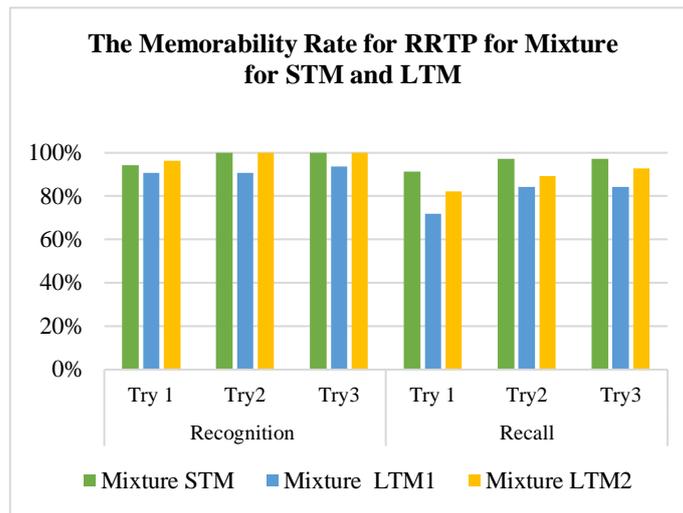

Figure 8. The successful login rate for RRTP for mixture words

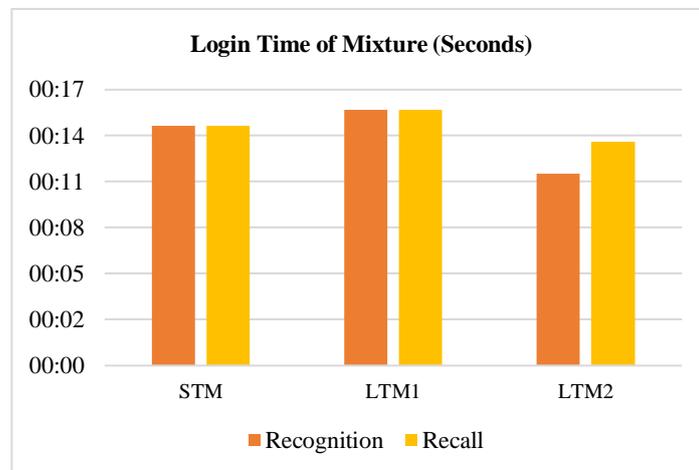

Figure 9. Login time for RRTP of mixture password

Table 9. Examples of some methods of RRTP for mixture password

| Examples of Password Creation Structures | | |
|---|---|---|
| | Phrase (Grammar flow) Participants (91.42 %) | Phrase (No grammar flow) Participants (8.57 %) |
| Mixture | -Internet and smartphone are important<br><br>-Snapchat friend is unsafe<br><br>- president is powerful and useful | -google become but dangerous<br><br>-Instagram applicable easy use<br><br>- computer learn is dislike |

Table 10. The enhancement on memorability rate after a reminder mixture password

| Reminder (Mixture) | | | |
|---|---|---|---|
| Recognition | | Recall | |
| No. Participants | Correctness | No. Participants | Correctness |
| First Reminder (0) | 0 | First Reminder (1) | 0 |
| Second Reminder (2) | 2 | Second Reminder (5) | 4 |

## 5. DISCUSSION

The memorability rate average for recognition and recall nouns is 74.97 % in LTM. The main reasons for the memorability shortage of nouns passwords belong to random selection or password creation structures. The memorability of nouns recognition is influenced by another factor which is 37.5 % of two chunks users have same words category of their password in displayed words set which confused them to retrieve the correct passwords such as ("morning" and "night") or ("YouTube" and "Snapchat"). Additionally, the words with the same first letter and close pronunciation, such as "story" and "store," made a difference in retrieving the correct password. The structures of password creation for recognition and recall nouns play an important role in retrieving the correct password. In contrast, the average memorability rate of recognition and recall of mixture passwords is 89.68 % in LTM. The slight decrease in the memorability rate for mixture recognition and recall passwords was caused by no grammar flow or spelling mistakes. Organizing the word set of nouns recognition passwords is very important as well as, the structure of password creation for both nouns and mixture words has a role in memorability. Overall, the memorability average for mixture words is higher than nouns for both conditions (**H5** supported). Some participants of nouns built a relationship between recognition and recall passwords that aided them to retrieve the correct passwords for both conditions in LTM, as shown in table 11.

Table 11. The Relationships the participants applied between RRTP for nouns

| | Relationship Between Recognition and Recall | | |
|---|---|---|---|
| | **Type of Relationship** | **No. Participant** | **Example** |
| Nouns | Synonyms/antonyms | 1 | **Recognition** (student home hour day) <br> **Recall** (teacher house minute night) |
| | Body parts and sides | 1 | **Recognition** (head eye back end) <br> **Recall** (arm hand front side) |
| | Antonyms | 1 | **Recognition** (government right city WhatsApp) <br> **Recall** (president left state Twitter) |

## 6. LIMITATIONS AND FUTURE WORK

The limitation of this study is the long period which led to missing some participants. The words of recognition password influenced the memorability rate because the users' had the same words category in words set such as "Youtube" and "Facebook". The future work will be based on analyzing the words' sets of recognition conditions to enhance memorability. Also, enhance the security of the proposed system by applying some techniques that mitigate different attacks.

## 7. CONCLUSION

Using English nouns and mixture words as recognition and recall passwords have a role in enhancing human memory to retrieve the password easily. There is a main factor that play a role in the memorability rate for both nouns and mixture words which are the structures of password creation. The result shows a gap in memorability rate between nouns and mixture words for both conditions in LTM. The RRTP of nouns needs a psychological method that enhances its memorability rate for LTM.

International Journal of Network Security & Its Applications (IJNSA) Vol.14, No.2, March 2022

## AUTHORS

**Hassan M. Wasfi**, Ph.D. student in the Department of HCI at Iowa State University. I received my master's degree in computer security from Newcastle university in the United Kingdom in 2011. I received my bachelor's degree in computer science from King Abdul Aziz University in 2008. I worked at King Abdul Aziz university as a lecturer from 2014 to 2020 in the information technology department. My current research focuses primarily on the area of computer security, especially in the authentication system and its relationship with behavioral aspects of password creation for recognition and recalling textual passwords.

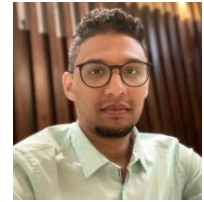

**Richard T. Stone**, Ph.D. is an Associate Professor in the Department of Industrial and Manufacturing Systems Engineering at Iowa State University. He received his Ph.D. in Industrial Engineering from the State University of New York at Buffalo in 2008. He also has an MS in Information Technology, a BS in Management Information Systems as well as university certificates in Robotics and Environmental Management Science. His current research focuses primarily on the area of human performance engineering, particularly applied biomedical, biomechanical, and cognitive engineering. Dr. Stone focuses on the human aspect of work across a wide range of domains (from welding to surgical operations and many things in between). Dr. Stone works extensively in tool, exoskeleton, and telerobotics technologies, as well as classic ergonomic evaluation.

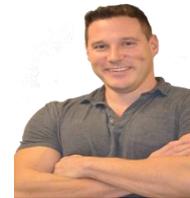